\documentclass[twocolumn,amsmath,amssymb,superscriptaddress]{revtex4}

\newcommand{\ket}[1]{|#1\rangle}             
\newcommand{\qo}[1]{``#1''}                  
\usepackage{graphicx}
\usepackage{amsfonts}

\begin{document}

\title{Liquid crystal spatial-mode converters for the orbital angular momentum of light}

\author{Sergei Slussarenko}
\affiliation{Dipartimento di Scienze Fisiche, Universit\`{a} di Napoli \qo{Federico II}, Complesso Universitario di Monte S. Angelo, 80126 Napoli, Italy}%
\email{sergei.slussarenko@na.infn.it}
\author{Bruno Piccirillo}%
\affiliation{Dipartimento di Scienze Fisiche, Universit\`{a} di Napoli \qo{Federico II}, Complesso Universitario di Monte S. Angelo, 80126 Napoli, Italy}%
\author{Vladimir Chigrinov}%
\affiliation{Hong Kong University of Science and Technology, Clear Water Bay, Kowloon, Hong Kong}%
\author{Lorenzo Marrucci}%
\affiliation{Dipartimento di Scienze Fisiche, Universit\`{a} di Napoli \qo{Federico II}, Complesso Universitario di Monte S. Angelo, 80126 Napoli, Italy}%
\affiliation{CNR-SPIN, Complesso Universitario di Monte S. Angelo, 80126 Napoli, Italy}%
\author{Enrico Santamato}%
\affiliation{Dipartimento di Scienze Fisiche, Universit\`{a} di Napoli \qo{Federico II}, Complesso Universitario di Monte S. Angelo, 80126 Napoli, Italy}%

\begin{abstract}
We present a tunable liquid crystal device that converts pure orbital angular momentum eigenmodes of a light beam into equal-weight superpositions of opposite-handed eigenmodes and vice versa. For specific input states, the device may thus simulate the behavior of a $\pi/2$ phase retarder in a given two-dimensional orbital angular momentum subspace, analogous to a quarter-wave plate for optical polarization. A variant of the same device generates the same final modes starting from Gaussian input.
\end{abstract}

\maketitle

\section{\label{sec:intro}Introduction}
During the last twenty years, the orbital angular momentum of light (OAM) has attracted a growing attention in the scientific world and is finding more and more applications in various fields of optics~\cite{rev_yao11}. It is particularly interesting for quantum information applications since, in contrast to the two-dimensional (2D) photon spin angular momentum space (SAM), the OAM space is inherently multidimensional and thus can be used for the implementation (separately, or together with the spin space) of photonic ``qudits'', i.e. multilevel quantum states used as elemental information carriers~\cite{molina04,nagali10a,nagali10}, and of decoherence-free photonic qubits, e.g. for alignment-free quantum communication~\cite{dambrosio12natcomm}.

The eigenstates of the Hilbert space associated to SAM correspond to left and right circular polarization states, here denoted as $\ket{L}_{p}$ and $\ket{R}_{p}$, respectively. OAM Hilbert space eigenstates, denoted as $\ket{\ell}_{o}$, correspond to a paraxial beam having an azimuthal phase dependence $\exp{(i\ell\varphi)}$, with integer $\ell$, where $\varphi$ is the azimuthal angle around the beam axis. While the main interest in OAM ultimately arises from its higher dimensionality~\cite{romero12}, a 2D OAM subspace is often adopted to implement qubit-based quantum information protocols or fundamental tests of quantum mechanics~\cite{jack10,romero10}. In these cases a close analogy with the spin Hilbert space can be made, mapping the circular-polarization states $\ket{L}_{p}$ and $\ket{R}_{p}$ onto OAM eigenstates $\ket{\pm\ell}_{o}$ and linear-polarization states $\ket{\theta}_{p}=\frac{1}{\sqrt{2}}(\ket{L}_{p}+e^{i\theta}\ket{R}_{p})$ onto superposition states $\ket{\theta_\ell}_{o}=\frac{1}{\sqrt{2}}(\ket{\ell}_{o}+e^{i\theta}\ket{-\ell}_{o})$, where $\theta/2$ is the angle between the polarization orientation and a fixed reference axis \cite{padgett99,karimi10}.

\begin{figure}
\includegraphics[width=0.45\textwidth]{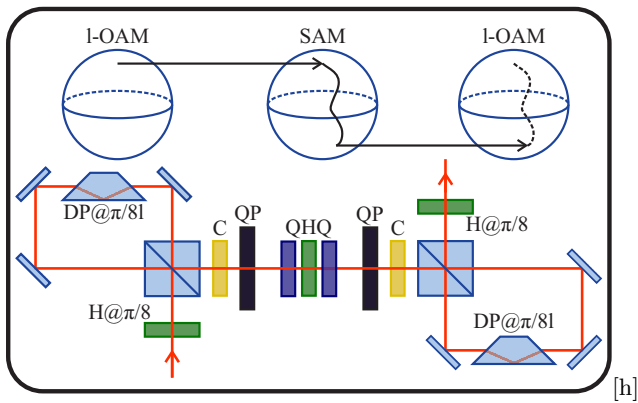}[h]
\caption{\label{fig:sto_setup} Bottom: OAM-space arbitrary state transformer setup scheme based on SAM-OAM transferrer devices. QP -- q-plate, H -- half-wave plate, Q -- quarter-wave plate, DP -- Dove prism, C -- Phase compensator. QHQ represents a unitary SAM gate. The quantity given after the @ symbol is the azimuthal angle at which the device axis must be oriented (in radians). Top: schematic representation of the setup action in terms of SAM and OAM Poincar{\'e} spheres.}
\end{figure}

While the polarization state of the photon can be easily controlled by polarizers, wave-plates and electro-optical phase retarders, for the OAM space there are not yet such convenient devices. Various techniques of OAM state generation have been introduced up to now, such as holograms~\cite{vasnet90,vasnet92} (including reconfigurable ones, made using spatial light modulators), spiral phase plates~\cite{beijersbergen94}, q-plates~\cite{marrucci06,marrucci06a}, various interferometric setups~\cite{leach02,slussarenko10oe}, and others. Several setups were proposed and demonstrated for the manipulation of a hybrid spin-orbit 4D space, where the 2D OAM space is used along with 2D SAM space~\cite{slussarenko10oe,slussarenko09}. Fewer are the existing methods for manipulating the already generated pure OAM states. To achieve full control of a two-dimensional quantum state, that is for being able to perform an arbitrary path on the Poincar{\'e} (or Bloch) sphere that represents the 2D Hilbert space, both a $\pi$ and a $\pi/2$ phase-retarders are required~\cite{bhandari97}. In SAM space such tools are provided by half- and quarter-wave birefringent retardation plates, respectively.  In case of 2D OAM space, $\pi$ phase retarders changing $\ket{\ell}_{o}$ into $\ket{-\ell}_{o}$ and $\pi/2$ phase retarders for $\ell=1$ space are already available, e.g.\ image rotation devices such as Dove prisms and cylindrical-lens $\pi$-converters~\cite{beijersbergen93}. Convenient OAM $\pi/2$ phase retarders, that change $\ket{\ell}_{o}$ into $\ket{\theta_\ell}_{o}$ for any $\ell$, are instead still missing. A currently feasible -- though rather complex -- scheme to this purpose may exploit the SAM-to-OAM (STO) state transferrers~\cite{karimi10,dambrosio12}, which can convert any polarization state $\alpha\ket{L}_{p}+\beta\ket{R}_{p}$ into a corresponding OAM state $\alpha\ket{\ell}_{o}+\beta\ket{-\ell}_{o}$ and vice versa. Such devices perform the transformation in a deterministic way with theoretical 100\% efficiency and are realized through a combination of a q-plate, which interfaces OAM and SAM spaces of the photon, and a polarizing Sagnac interferometer with a Dove prism (PSI)~\cite{slussarenko10oe}. Therefore, in order to implement the desired OAM state transformation, a transferrer should be first used to transfer the OAM state into the SAM space. Then, a quarter-wave plate induces the desired $\pi/2$ phase transformation and, finally, a second transferrer is used to return to the OAM space. This approach is shown schematically in figure~\ref{fig:sto_setup}. The same scheme can be used to give rise to any other unitary transformation in the OAM space, as the quarter-wave plate can be replaced by a suitable set of wave plates (e.g., a half-wave plate, sandwiched between two quarter-wave plates) that together, can perform any unitary transformation in the SAM Hilbert space~\cite{bhandari97}. The realization of STO and STO-based unitary $\ell$-OAM gates is only possible due to the q-plates, that introduce a controlled coupling between SAM and OAM. A simplified probabilistic scheme, where PSIs are replaced by filtering devices (a polarizer or coupling into a single-mode optical fiber, depending on the transformation direction), can be used for the same goal, decreasing however, the efficiency of the whole apparatus to a theoretical maximum of 25\%~\cite{nagali09a}. It is worth noting that, by combining several STO-based interferometers with q-plates having different topological charges $q$, an OAM state manipulation beyond a single 2D subspace would also be possible.
In this paper, we demonstrate a liquid-crystal Pancharatnam-Berry optical element (PBOE) able to change in a single pass the OAM state of the beam from an eigenstate $\ket{\ell}_{o}$ into a superposition state $\ket{\theta_\ell}_{o}$ and vice versa, so as to partly mimic the behavior of a $\pi/2$ phase transformation in OAM space. Moreover, we present a similar PBOE device able to generate the $\ket{\theta_\ell}_{o}$ state directly from a Gaussian input.

\section{\label{sec:theory}Theoretical background}
\begin{figure}
\includegraphics[width=0.45\textwidth]{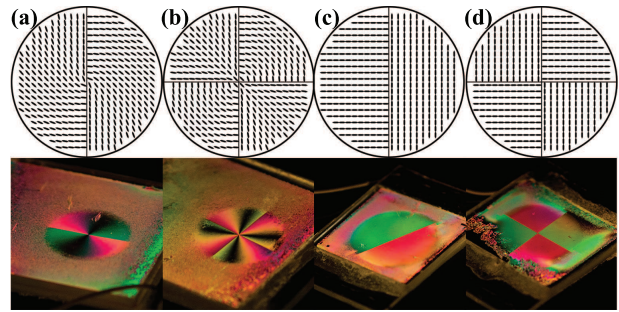}
\caption{\label{fig:converters} Top row: optical axis orientation patterns for (a) $\pi/2$ MC for $\ell=1$, (b) $\pi/2$ MC for $\ell=2$, (c) MG for $\ell=1$,  and (d) MG for $\ell=2$. Bottom row: corresponding fabricated samples seen between crossed polarizers under oblique illumination. Different colours correspond to different optical axis orientations, with dark areas corresponding to the zones where the optical axis is oriented parallel to one of the polarizers.}
\end{figure}

A Pancharatnam-Berry optical element is a space-variant phase element that exploits the geometrical phase arising in polarization manipulations for reshaping the optical wavefront~\cite{bhandari97,bomzon01a,marrucci06a}. This should not be confused with the analogous geometrical phase arising in the manipulation of OAM modes~\cite{vanenk93,galvez03}. Our PBOE is realized as a birefringent waveplate of uniform phase retardation $\delta$, whose angle $\alpha$ between slow optical axis and the $x$ axis of the fixed laboratory reference frame is not constant, but is described by a prescribed function $\alpha=f(\rho,\varphi)$, where $\rho$ and $\varphi$ are the polar coordinates in the waveplate transverse plane. It can be easily demonstrated with the Jones matrix approach, that the corresponding transformation matrix in the circular polarization basis $\{\ket{L}_{p},\ket{R}_{p}\}$ is given by
\begin{eqnarray}\label{eq:matrix}
T(\rho,\phi)=&\cos(\delta/2)\left(
                               \begin{array}{cc}
                                 1 & 0 \\
                                 0 & 1 \\
                               \end{array}
                             \right)  +\nonumber\\&i\sin(\delta/2)\left(
                                                                     \begin{array}{cc}
                                                                       0 & e^{-2i f(\rho,\varphi)} \\
                                                                       e^{2i f(\rho,\varphi)} & 0 \\
                                                                     \end{array}
                                                                   \right).
\end{eqnarray}
In other words, a fraction of the incident circularly-polarized light undergoes a helicity inversion and gains an additional phase factor of $\pm2f(\rho,\varphi)$, where the sign depends on the input polarization helicity. This additional phase factor is not due to refractive index change or non-uniform thickness of the optical element, but to the geometrical phase arising from the space-variant polarization manipulation. The efficiency of the beam conversion depends on the phase retardation $\delta$ and is maximum for the half-wave $\delta=\pi$ condition. To calculate the pattern $f(\rho,\varphi)$ needed to perform a desired state transformation, one should use the phase difference between the output and input beam states. To transform the eingenstate $\ket{\ell_1}_{o}$ into an equal-weight superposition $\ket{\theta_{\ell}}_{o}$ (with $\theta=0$) and vice versa the corresponding pattern is given by:
\begin{equation}\label{eq:charge}
f_c(\varphi,\ell,\ell_1)=\frac{1}{2}(-\ell_1\varphi + Arg(e^{i\ell\varphi}+e^{-i\ell\varphi})).
\end{equation}
Setting $\ell_1=\pm\ell$ yields a $\pi/2$ mode converter (MC) device, while setting $\ell_1=0$ and arbitrary $\ell$ yields a mode generator (MG) device that can generate a $\ket{\theta_\ell}_{o}$ state from the Gaussian $\ell=0$ input directly. Setting $\ell\neq\ell_1$ or tuning the phase retardation of the device (and thus its conversion efficiency) allows more complex state manipulations even beyond the original 2D OAM subspace. Some examples of these optical axis patterns are given in figure~\ref{fig:converters}, together with images of corresponding fabricated samples.

The phase factor $\theta$ of the output state $\ket{\theta_\ell}_{o}$ can be controlled by a physical rotation of the device and is given by $\theta=2\ell\xi +\pi$, where $\xi$ is the angle between one of the $\pi/2$ jump lines of the device pattern and the $x$-axis of the fixed reference frame. Since the phase transformation is polarization dependent, the correct output, in general, will be obtained only for one circular input polarization. Nevertheless, the MGs will give the same output, up to a global phase factor, for any input polarization. In the case of the MCs, the correct converted output $\ket{\theta_\ell}_{o}$ will be obtained only for $\ket{L,\ell}$ and $\ket{R,-\ell}$ input spin-orbit states. For the other two basis states $\ket{L,-\ell}$ and $\ket{R,\ell}$ of the spin-orbit combined space, the converted state will not correspond to a combination of $\ket{\pm\ell}_{o}$ states and will be brought outside the initial 2D Hilbert space. In the case of the inverse transformation, one obtains $\ket{\ell}_{o}$ as the output state for a $\ket{L,\theta_\ell}$ input, and $\ket{-\ell}_{o}$ for the $\ket{R,\theta_\ell}$ input. This polarization dependence is a limitation of our MC device, as at fixed input polarization it mimics the $\pi/2$ phase transformation in the OAM 2D subspace only for specific input states. Yet, the ease of use and compactness of our device makes it still interesting and suitable for certain OAM manipulation applications.

\section{\label{sec:experiment}Fabrication and characterization}
\begin{figure}
\includegraphics[width=0.45\textwidth]{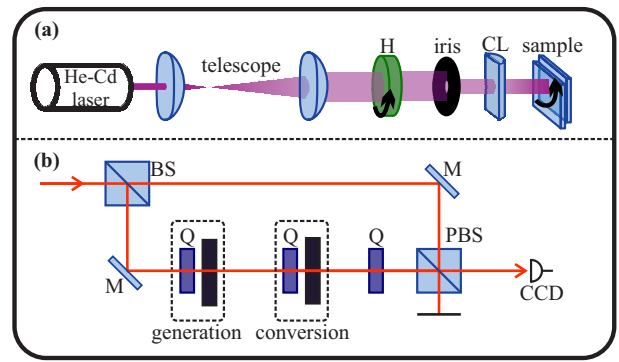}
\caption{\label{fig:fab_setup} (a) Liquid crystal PBOE fabrication setup scheme. H -- Half-wave plate, CL -- cylindrical lens. (b) Mach-Zehnder interferometer setup used to analyze the fabricated samples. Q -- quarter-wave plate, BS -- beam splitter, M - mirror, PBS -- polarizing beam splitter, CCD -- CCD camera. The "generation" and "conversion" blocks represent different combinations of quater-wave plate and various PBOEs, such as q-plates, MGs and MCs that were used for both generation of the desired input state and its subsequent conversion by the tested device.}
\end{figure}

We used nematic liquid crystals (LC) to fabricate our PBOE devices. The desired planar alignment of the LC was induced using a photoalignment technique~\cite{book_chigrinov_PA}. The scheme of our fabrication setup is shown in figure~\ref{fig:fab_setup}(a). The sample was made from two glass substrates, spin-coated with $1\%$ solution of sulphonic azo-dye SD1 (Dainippon Ink and Chemicals) in dimethylformamide (DMF) for $30$~s at $3000$~rpm. We used glass windows with conducting Indium-Tin-Oxide (ITO) coating so to have the possibility of applying an external electrical field to the LC film. After the evaporation of the solvent, by soft-baking at $120~\ensuremath{^\circ}C$ for $5$~min, the glasses were assembled together and $6~{\mu}$m glass spacers were used to define the cell gap. The SD1 surfactant provides planar alignment for the LC in the direction perpendicular to the incident light polarization, with anchoring energy comparable with the polyimide rubbing based alignment~\cite{book_chigrinov_PA}.  A He-Cd 325~nm laser of 10~mW power (from VM-TIM) was used as collimated UV light source. The polarized laser beam was expanded by a set of the two confocal lenses, sent through a half-wave plate and focused on a sample with a cylindrical lens of 75~mm focal length. Both the waveplate and the sample were attached to rotating mounts controlled by computer through step-motors. By programming the relative step size of two motorized mounts it is possible to impress any orientation pattern with an angular dependence. After the exposure, the cell was filled with nematic liquid crystal (E7 from Merck) and sealed with epoxy glue. A total exposure of 1.5~h was enough to provide high quality LC alignment. PBOE devices working with $|\ell|=1$, $|\ell|=2$ (see figure~\ref{fig:converters}) and others were realized in this work.

As in the case of q-plates, or other LC cells, the birefringence, and thus the phase retardation $\delta$ appearing in~\ref{eq:matrix} was controlled by applying an external AC electric field. Typical maximum and minimum efficiencies, switching times and overall tuning behavior of our devices were found to be identical to those of q-plates and have been reported elsewhere~\cite{piccirillo10,slussarenko10}.

As a main test of our PBOEs, we studied the wavefront transformation of a laser beam passing through the sample. To this purpose, we assembled a Mach-Zehnder interferometer (MZI), shown in figure~\ref{fig:fab_setup}(b). In one of the arms of the MZI the input linearly polarized beam was manipulated by means of quater-wave plates, q-plates, MGs and MCs. The other arm was used as reference $TEM_{00}$ beam. The interference patterns were then recorded by a CCD camera (Hamamatsu C5405). By blocking the reference arm it was possible to record the intensity pattern of the analyzed beam, too.

\begin{figure}
\includegraphics[width=0.45\textwidth]{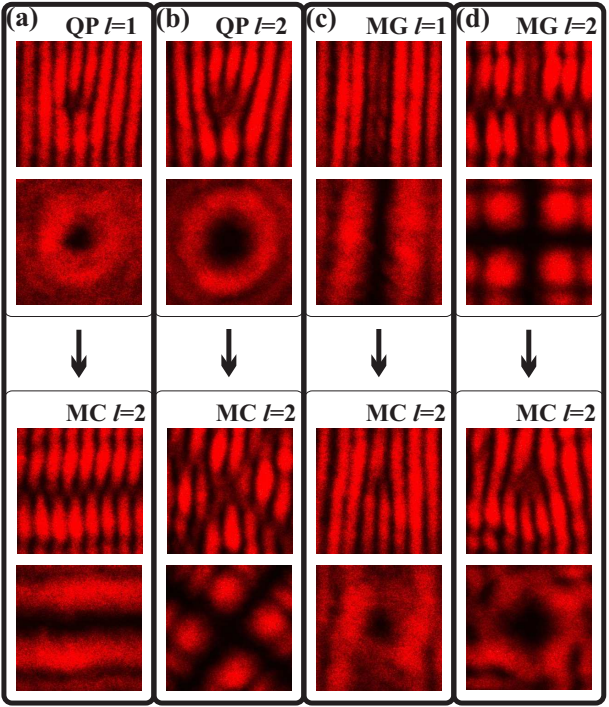}
\caption{\label{fig:conversions_lg_hg} Intensity and interference images for four different configurations of "generation" (top section) and "conversion" (bottom section) blocks shown in figure~\ref{fig:fab_setup}(b). (a) -- q-plate for $\ell=1$ as generator and MC for $\ell=1$ as converter, (b) -- q-plate for $\ell=2$ as generator and MC for $\ell=2$ as converter, (c) -- MG for $\ell=1$ as generator and MC for $\ell=1$ as inverse converter,  (d) -- MG for $\ell=2$ as generator and MC for $\ell=2$ as inverse converter.}
\end{figure}

In case of the $\ket{\ell}_{o}$ state, an optical vortex for the intensity distribution and a fork-like grating pattern with a fringe disclination at the vortex location are expected. In case of $\ket{\theta_\ell}_{o}$ states, a set of $2\ell$ light lobes or a set of $2\ell$ sectors with gratings shifted by a half-period with respect to the adjacent ones are expected for the intensity and phase distributions, respectively.

For the MCs characterization we used circularly polarized TEM$_{00}$ beam and a q-plate with $q=0.5$ and $q=1$ to prepare $\ket{\ell}_{o}$ modes with $\ell=1$ and $2$, respectively. The generated modes were then transformed by a corresponding MC converter into $\ket{\theta_\ell}_{o}$ modes that have 2 and 4 intensity lobes, respectively. The registered $\ket{\ell}_{o}$ input and $\ket{\theta_\ell}_{o}$ output interference and intensity images are shown in figure~\ref{fig:conversions_lg_hg}(a) for $\ell=1$ and (b) for $\ell=2$. For the MGs characterization, the q-plate was replaced by a MG for $\ell=1$ and $2$, so as to generate a $\ket{\theta_\ell}_{o}$ beam from a TEM$_{00}$ input directly. Then the inverse $\ket{\theta_\ell}_{o}\rightarrow\ket{-\ell}_{o}$ transformation was performed with the corresponding MC converters. The recorded intensity and interference images for the generated input and converted states are shown in figure~\ref{fig:conversions_lg_hg}(c) for $\ell=1$ and (d) for $\ell=2$.

\begin{figure}
\includegraphics[width=0.45\textwidth]{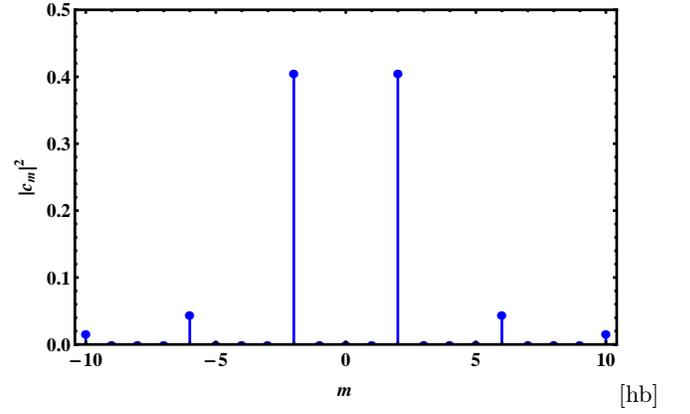}[hb]
\caption{\label{fig:spectrum} Power spiral spectrum of an output from the MC with $\ell=2$.}
\end{figure}

Our converters, in the same way as binary holograms or any other phase-only transforming device, can only change the phase of the input beam. In this case, the angular dependence of the output state after the MC will have the phase factor $\exp({i Arg(e^{i\ell\varphi}+e^{-i\theta}e^{-i\ell\varphi})})$ which is different from the desired amplitude factor $(e^{i\ell\varphi}+e^{-i\theta}e^{-i\ell\varphi})/\sqrt{2}$~\cite{jack08,jack09}. A straightforward calculation shows that such state does not correspond to the sum of the two eigenmodes with helical phases $\exp({\pm i \ell \varphi})$, but to an infinite sum of helical modes $\exp({i m \varphi})$ with intensity coefficients given by
\begin{eqnarray}\label{eq:spectra}
|C(\ell,m)|^2 = \left\{ \begin{array}{ll}
4\ell^2/({\pi^2 m^2}) & m=(2k+1)\ell,~~k~\epsilon~\mathbb{Z}\\
0 & m\neq(2k+1)\ell,~~k~\epsilon~\mathbb{Z}
\end{array} \right.
\end{eqnarray}
An example of $|C(\ell,m)|^2$ for $\ell=2$ is shown in figure~\ref{fig:spectrum}.
As it can be calculated from~\ref{eq:spectra}, over 80\% of the output intensity belongs to the subspace of interest $\ket{\pm\ell}_{o}$. The remaining intensity is distributed among higher OAM orders where non-zero coefficients correspond to $m=(2k+1)\ell$ (where $k$ is integer). While the $\ell$ value defines what modes are present in the spectrum, the intensity distribution among these non-zero high order components does not depend on $\ell$. For example, the ratio between the intensity of the nearest higher mode $m=3\ell$ and the mode $m=\ell$ is equal to $1/9$. In terms of efficiency and state fidelity, such higher modes are usually treated as intensity or photon losses that do not influence the state fidelity within the subspace $\{\ket{\ell}_{o},\ket{-\ell}_{o}\}$ and are usually filtered out at the detection phase.

\begin{figure}
\includegraphics[width=0.45\textwidth]{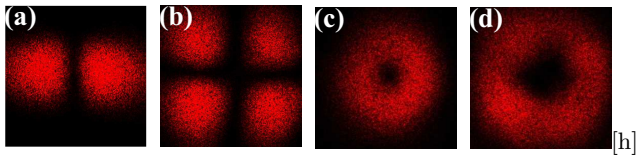}[h]
\caption{\label{fig:farfield_lg_hg} Far field intensity images of output beams after analyzed PBOEs. (a) and (b) --  MG for $\ell=1$ and $\ell=2$ that correspond to near field distributions shown in figure~\ref{fig:conversions_lg_hg}(c) and (d) top rows, respectively. (c) and (d) -- for inverse conversions by MC for $\ell=1$ and $\ell=2$ that correspond to near field distributions shown in figure~\ref{fig:conversions_lg_hg}(c) and (d) bottom rows, respectively.}
\end{figure}

Another feature of phase-only devices is in the way they affect the radial intensity distribution. If the input beam state is described in terms of Laguerre-Gauss modes $LG_{\ell,p}$ (where $\ell$ and $p$ are the azimuthal and radial indices, respectively), the phase-only transformations such as those induced by our PBOEs do not preserve the $p$ number, creating an output that has an intensity profile that varies during propagation below the Rayleigh range~\cite{karimi07,karimi09ol}. Indeed, as it can be seen, the intensity structure of the output beam in figure~\ref{fig:conversions_lg_hg}(c) and (d) (bottom section), even if an optical vortex is clearly visible, does not have the typical \qo{doughnut} shape. This is due both to the excitation of several $p$ modes and to the presence of additional higher-order $\ell$ modes, as mentioned above. An additional image was taken in the far-field of the output beam, by placing a lens and a microscope objective at the lens back focal plane. In the far-field, owing to a spatial filtering of all higher $p$ modes, the converted $\ket{\theta_\ell}_{o}$ beams appear to be made by a single set of $2\ell$ lobes around the beam axis, as shown in in figure~\ref{fig:farfield_lg_hg}(a) and (b) and$\ket{\ell}_{o}$ beams have clear optical vortex shapes with a single ring, as shown in figure~\ref{fig:farfield_lg_hg}(c) and (d).
\section{\label{sec:outro}Conclusions}
In conclusion, we have introduced a liquid crystal optical element (MC) that can perform a $\ket{\ell}_{o}\rightarrow\ket{\theta_\ell}_{o}$ operation on the beam OAM state or its inverse, thus mimicking the behavior of a $\pi/2$-phase shifter for specific input states. Moreover, we have also introduced a LC element (MG) that can be used for generating a $\ket{\theta_\ell}_{o}$ state directly from a TEM$_{00}$ gaussian input. Both devices operate by introducing a polarization-dependent phase-only transformation of the input beam. It is worth noting that the devices introduced here can be also used in the single photon quantum regime. We think that the MC and MG devices may find application in all fields in which a complete manipulation of the light OAM is needed, such as, for instance, optical communication, optical computing, and quantum cryptography.

\begin{acknowledgments}
This work was supported by the Future and Emerging Technologies (FET) program within the Seventh Framework Programme for Research of the European Commission, under FET-Open Grant No.\ 255914, PHORBITECH and HKUST grant CERG 612310.
\end{acknowledgments}
\section*{References}
\bibliography{main}
\bibliographystyle{unsrt}

\end{document}